\documentclass[pra,superscriptaddress,nofootinbib,noshowpacs,preprintnumbers,longbibliography,floatfix,twocolumn]{revtex4-2}
\usepackage[utf8]{inputenc}
\usepackage{graphicx}
\usepackage{float}
\usepackage{amssymb}
\usepackage{amsmath}  
\usepackage{adjustbox}
\usepackage{mathtools}
\usepackage{dsfont}
\usepackage{array}
\usepackage{bm,fixmath}
\usepackage{mathrsfs}
\usepackage{pifont}
\usepackage{multirow}
\usepackage{upgreek}
\usepackage{xcolor}
\usepackage{bm}
\usepackage{bbm}
\usepackage{physics}
\usepackage{comment}
\usepackage{tikz}
\usetikzlibrary{quantikz}
\usepackage[paperwidth=240mm,paperheight=297mm,centering,hmargin=3cm,vmargin=3cm]{geometry}
\usepackage[compat=0.6]{yquant}
\yquantdefinebox{dots}[inner sep=0pt]{$\dots$}
\usepackage[colorlinks = true,
            linkcolor = blue,
            urlcolor  = blue,
            citecolor = violet,
            anchorcolor = blue]{hyperref}

\usepackage{orcidlink} 

\begin{document}

\title{Quantum Brush: A quantum computing-based tool for digital painting}

\author{Jo\~ao~S.~Ferreira \orcidlink{0000-0003-1054-9518}}
\email{joao@mothquantum.com}
\affiliation{Moth, Arlesheim, BL, Switzerland}

\author{Arianna~Crippa \orcidlink{0000-0003-2376-5682}}
\affiliation{Deutsches Elektronen-Synchrotron DESY, Platanenallee 6, 15738 Zeuthen, Germany
}

\author{Astryd~Park~\orcidlink{0009-0002-9367-0902}}
\affiliation{
    Moth, Somerset House WC2R 1LA, London, United Kingdom
}

\author{Daniel~Bultrini}
\affiliation{Moth, Arlesheim, BL, Switzerland}

\author{Pierre Fromholz}
\affiliation{Moth, Arlesheim, BL, Switzerland}

\author{Roman Lipski}
\affiliation{Atelier Roman Lipski, Wiener Str. 10, 10999 Berlin, Germany}

\author{Karl~Jansen~\orcidlink{0000-0002-1574-7591}}
\affiliation{Deutsches Elektronen-Synchrotron DESY, Platanenallee 6, 15738 Zeuthen, Germany
}
\affiliation{
 Computation-Based Science and Technology Research Center, The Cyprus Institute, 20 Kavafi Street,
2121 Nicosia, Cyprus
}

\author{James R. Wootton}
\affiliation{Moth, Arlesheim, BL, Switzerland}

\date{\today}

\begin{abstract}
We present Quantum Brush, an open-source digital painting tool that harnesses quantum computing to generate novel artistic expressions. The tool includes four different brushes that translate strokes into unique quantum algorithms, each highlighting a different way in which quantum effects can produce novel aesthetics. Each brush is designed to be compatible with the current noisy intermediate-scale quantum (NISQ) devices, as demonstrated by executing them on IQM's \texttt{Sirius} device.
\end{abstract}
\maketitle

\section{Introduction}

Throughout history, technological advances have expanded artistic possibilities and birthed novel aesthetics through the exploration of new tools, methods, and workflows, including oil paints, photography, computer graphics, and machine learning. As quantum technologies begin to enter the mainstream, a natural and provocative question arises: \textit{Can there be such a thing as a quantum aesthetic or artistic style?}

Artists have already started to address this question~\cite{morales2003realismo}. A growing number of works across music~\cite{mirandaQuantumComputerMusic2022a,Clemente:2022qce,Itaborai:2023xha,Itaborai:2024fxj}, visual art~\cite{lioretQuantumArt2016,crippa2025quantumcomputinginspiredpaintings}, and interactive media~\cite{ferreira2025qrc} explore different aspects of modern quantum research. This emerging field, here referred to as \textit{quantum art}, has gained recognition through dedicated exhibitions, performances, and academic conferences~\cite{qiskitMakingInvisibleVisible2021,Quantumblur,baumgarten_quantumjungle_2023,qartworkshop}. While a universally accepted definition of quantum aesthetics is still evolving, it typically involves the use of quantum principles, algorithms, or devices to create experiences that can be fundamentally different from classical means.
\begin{figure}
    \centering
    \includegraphics[width=\columnwidth]{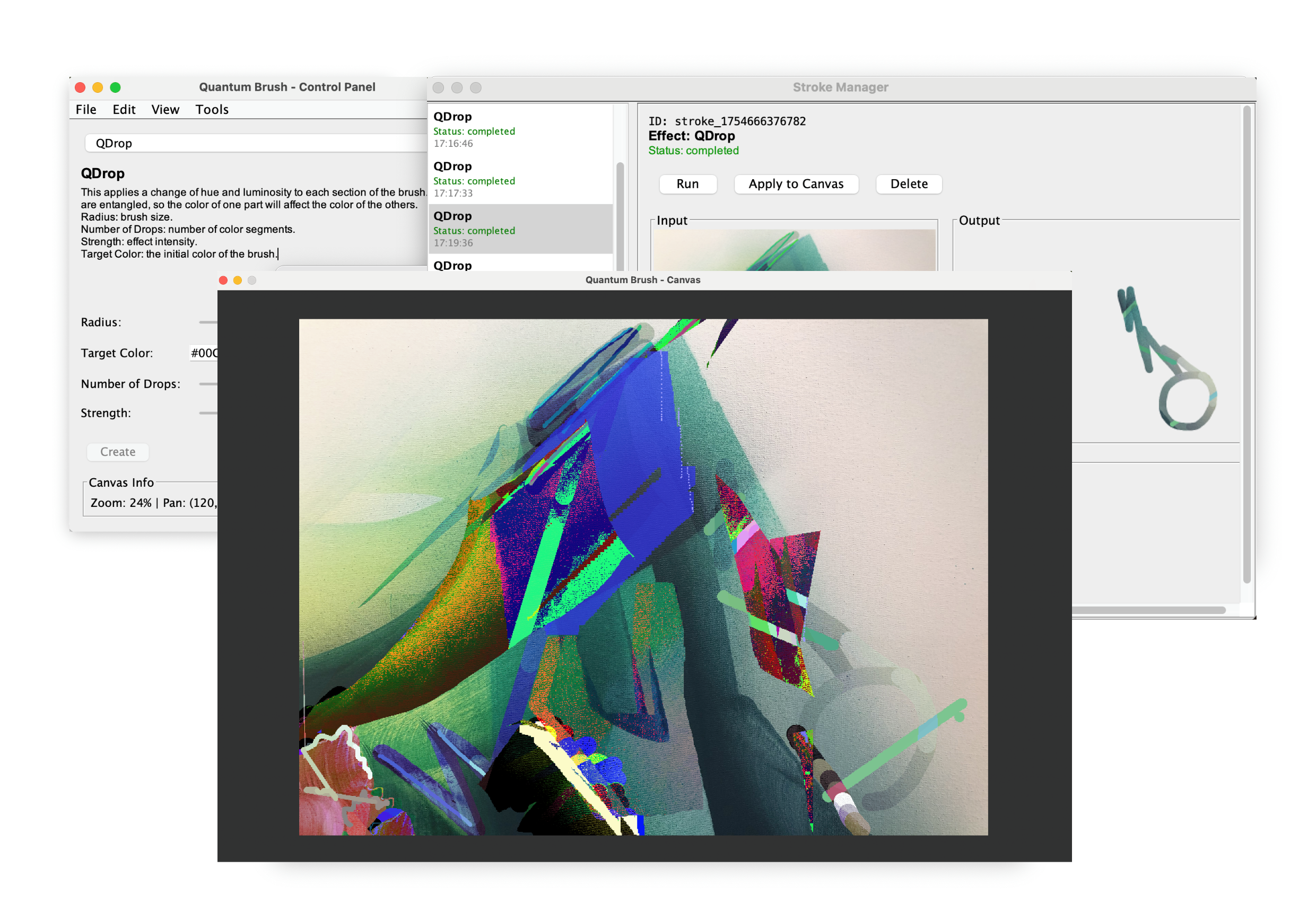}
    \caption{\textbf{Screenshots of the Quantum Brush application.} The stand alone open-source application allows users to import and modify images using different quantum brushes. Its workflow is specifically adapted to the constraints of quantum processing units, including slow execution times and run-to-run variability. \label{fig:app}}
\end{figure}
One of the key tools driving this creative potential is quantum computing. Quantum computers leverage the principles of quantum mechanics to process information in fundamentally non-classical ways. In contrast to classical computers, which operate on \textit{bits} that can be either \texttt{0} or \texttt{1}, 
quantum computers go beyond, and use quantum bits, or \textit{qubits}~\cite{nielsen2002quantum,wong2022introduction}. Using a metaphor: if \texttt{0} would be north, and \texttt{1} would be south, then a qubit, like a compass needle is able to explore many other possible states in-between. All states that are not simply \texttt{0} or \texttt{1} are called superpositions. With multiple qubits, the interaction of these superposition states becomes possible; in this case, we say that the qubits are entangled. Qubits in quantum computers typically start in the \texttt{0} state, with superpositions and entangled states obtained by manipulating the qubits through sequences of logical \textit{quantum gates}.

Through the creation and manipulation of superpositions, the state of qubits becomes subject to wave-like interference effects and non-classical correlations. Harnessing these effects enables the design of unique quantum algorithms for quantum art.
Yet, at the end of a quantum process, we must extract an output. Through this process, known as measurement, any remaining superpositions will `collapse' into a random choice between \texttt{0}s and \texttt{1}s, with the superposition determining the probability of each set of outcomes.
Highlighting the underlying quantum structures from the measurement outcome remains one of the central challenges for any quantum art tool.

In this work, we contribute to this growing intersection of quantum science and artistic expression by introducing the \textbf{Quantum Brush}, a digital painting tool whose behaviour is driven by quantum algorithms. As artists interact with the canvas using familiar brush-like gestures, the tool constructs and executes quantum algorithms in real time, see Fig.~\ref{fig:app}. The resulting quantum state is measured and mapped back onto the canvas, modulating colour and form in ways that reflect the underlying invisible quantum dynamics. Our approach opens a new avenue in quantum generative art, where aesthetic outcomes are not only inspired by but also shaped through quantum mechanics.

The Quantum Brush is released as a standalone, open-source application~\cite{quantumbrush_repo}. At the time of the writing, it includes 4 distinct brushes, discussed in Sec.~\ref{sec:qbrushes}, which explore different dimensions of our idea of quantum art: Aquarela (Sec.~\ref{sec:Aquarela}), Heisenbrush
(Sec.~\ref{sec:heisenbrush}), Smudge (Sec.~\ref{sec:qdamp}) and Collage (Sec.~\ref{sec:ctrlq}). The artistic intention and scientific implementation are explained in each case.
Sec.~\ref{sec:usage} provides an overview of the user interface and some details about the how the application functions. In Sec.~\ref{sec:artwork}, we present the experience of one of the authors (R.L.) when creating a new artwork using the quantum brush. Finally, Sec.~\ref{sec:discussion} discusses the broader implications of this tool for both the quantum and artistic communities.

\section{Quantum Brushes}\label{sec:qbrushes}

In this section, we introduce four distinct quantum brushes, each designed to explore a different facet of quantum aesthetics. The design of these brushes was guided by a set of soft guidelines aimed at ensuring both artistic usability and physical feasibility:

\begin{itemize}
    \item The brushes must act \textit{locally} on the image, affecting only pixels along the stroke's path.
    \item Each brush must feature a \textit{continuous control parameter} such that, when set to zero, the quantum effect vanishes (aside from intrinsic noise, i.e. the uncontrolled influence of the environment on the quantum hardware).
    \item The algorithms must be executable on current, noisy intermediate-scale quantum (NISQ) hardware, avoiding reliance on fault-tolerant techniques.
    \item The algorithms must admit a scaling regime where classical simulation becomes intractable, requiring a true quantum processing unit (QPU).
\end{itemize}
Each brush has its own style and origin and may not satisfy all the constraints. 

For each brush, we present the underlying artistic motivation and illustrate its effect through the reinterpretation of a well-known artwork. Since the creative touches of the Quantum Brush can benefit from the imperfection in NISQ quantum hardware, we showcase results from both a noiseless quantum simulator and from the execution on IQM's \texttt{Sirius} device without any error-mitigation. For readers with a deeper interest in quantum mechanics, the sub-sections \emph{Implementation} provide the necessary technical discussion to understand the underlying algorithm design and behaviour.

\begin{figure}
    \centering
    \includegraphics[width=0.48\columnwidth]{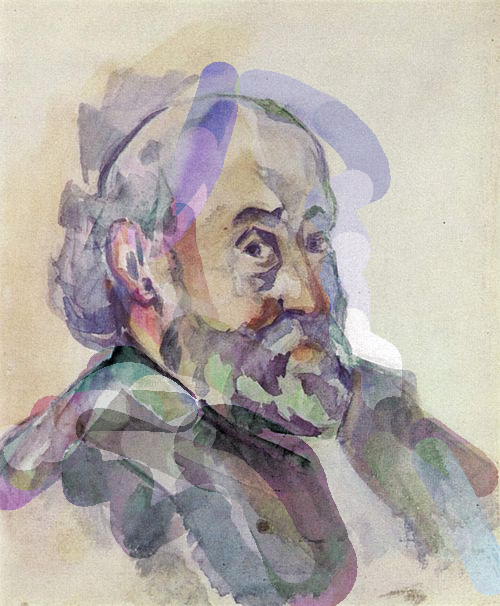}
    \includegraphics[width=0.48\columnwidth]{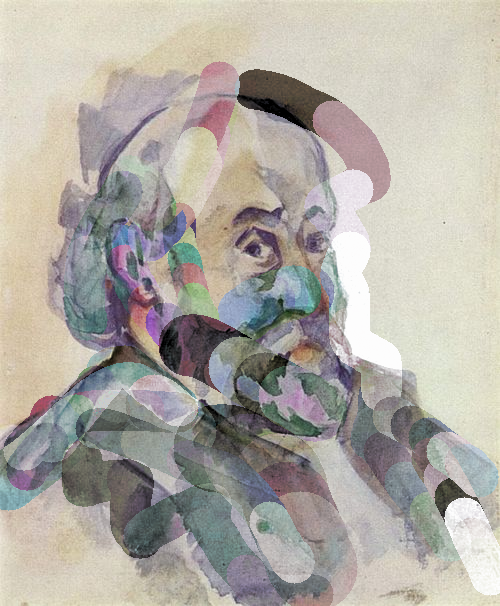}
    \caption{\textbf{Application of the Aquarela brush} on Paul Cézanne’s 1895 self-portrait~\cite{cezanne}, blending into the watercolour texture while exhibiting novel colour shifts. Left: Simulation on a noiseless backend. Right: Execution on IQM's \texttt{Sirius} device.}
    \label{fig:Aquarela}
\end{figure}

\begin{figure}
\centering
\begin{tikzpicture}
  \begin{yquant}
    qubit {$\ket{0}$} q;
    qubit {\ldots} n;
    qubit {$\ket{0}$} q[+1];
    qubit {$\ket{0}$} a;
   
    box {R$_y(\theta_1)$} q[0];
    box {R$_z(\phi_1)$} q[0] ;
    box {R$_y(\theta_N)$} q[1];
    box {R$_z(\phi_N)$} q[1] ;
        
    x a;
    output {$\ket{(\phi_1,\theta_1)}$} q[0];
    output {\ldots} n;
    output {$\ket{(\phi_N,\theta_N)}$} q[1];
    output {$\ket{1}$} a;
    
  \end{yquant}
\end{tikzpicture}
\caption{\textbf{Preparation phase of the Aquarela brush}. Each average colour of each N segments is encoded into its own qubit, and the brush qubit is initialized to $\vert 1\rangle$.}
\label{fig:Aquarelaembedding}
\end{figure}

\begin{figure*}[t]
\centering
\begin{tikzpicture}
\begin{yquant}
    qubit {$\ket{(\phi_{1},\theta_{1})}$} q;
    qubit {\ldots};
    qubit {$\ket{(\phi_{N},\theta_{N})}$} q[+1];
    qubit {$\ket{1}$} a;

    box {R$_z\left(-\gamma \phi_1\right)$} q[0] | a;
    box {R$_y\left(\gamma (\theta_B-\theta_1)\right)$} q[0] | a;
    box {R$_z\left(\gamma \phi_B\right)$} q[0] | a;
    box {R$_y\left(\frac{\pi}{3}\right)$} a ~ q[0];

    box {R$_z\left(-\gamma \phi_N\right)$} q[1] | a;
    box {R$_y\left(\gamma (\theta_B-\theta_N)\right)$} q[1] | a;
    box {R$_z\left(\gamma \phi_B\right)$} q[1] | a;
    box {R$_y\left(\frac{\pi}{3}\right)$} a ~ q[1];
  \end{yquant}
\end{tikzpicture}
\caption{\textbf{Diagram of the Aquarela circuit}. An ancilla qubit interacts with all qubits sequentially, making them rotate towards the colour of the brush. All qubits are measured at the end.}
\label{fig:Aquarela_circuit}
\end{figure*}
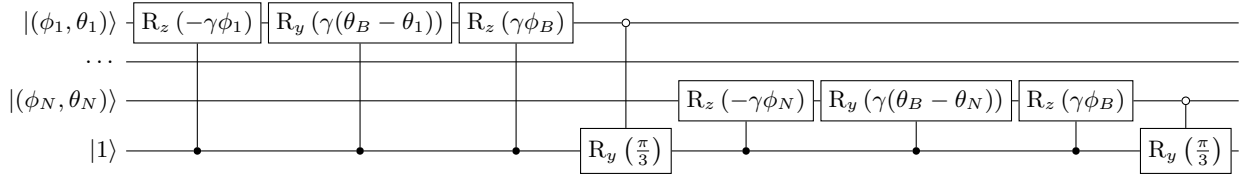

\subsection{Aquarela\label{sec:Aquarela}}
\subsubsection{Motivation}

The Aquarela brush is inspired by the translucent effect of a watercolour painting.
With Aquarela, the colours are not only blended, but \emph{entangled}. Each painted line becomes a quantum interaction, where the effect of the brush depends not only on its path but also on the state of the canvas beneath it. This process results in an evolution in which brush and canvas inform one another through quantum logic.

\subsubsection{Outcome}

To showcase the Aquarela brush, we applied it to a self-portrait by Paul Cézanne (1895), shown in Fig.~\ref{fig:Aquarela}, painted using watercolours.
We selected various tones of blue and purple for the brush, generating strokes of various sizes that weave into the texture of the painting while producing subtle changes in tones. These variations arise from the brush entangling all the colours along its path.

As expected, the presence of noise when running on real hardware leads to unique and unexpected results. Notice how a few face strokes converge to green and red tones not observed in the simulation. 
The added randomness from the noise is somewhat controlled as the tones always remain close to the simulated ones. 

\subsubsection{Implementation}

The user begins by selecting a desired colour for the brush and executes a continuous brush stroke on the canvas. This stroke is then divided into $N$ segments, with each segment corresponding to a single qubit. For each segment, we compute the mean HSL values of the underlying pixels. The hue (H) and luminosity (L) are then mapped to the spherical angles $\phi,\theta$ and encoded in the single-qubit state $\ket{(\phi,\theta)}$ through $R_{y,z}$ rotations along the $y,z$ axes shown in Fig.~\ref{fig:Aquarelaembedding}.
The auxiliary (brush) qubit is initialized in the state $\ket{1}$. The details of this encoding are described in App.~\ref{app:hsl}.

The brush qubit then interacts sequentially with each segment qubit, see Fig.~\ref{fig:Aquarela_circuit}. At each step, the brush attempts to \emph{steer} the segment qubit toward the brush's colour, encoded into $(\phi_B,\theta_B)$, using conditional rotations. The strength of the effect is controlled by $\gamma$, a user-controlled parameter.
Between each segment rotation, the brush qubit is also affected by the canvas through a negative controlled rotation of $\pi/3$ radians towards the $\ket{0}$ state (controlled on the qubit luminosity). The angle was chosen to yield the best artistic outcomes  with the constraint that it should not be close to $0$ or $\pi$. In the beginning of the stroke, an increasing $\ket{0}$ component leads to a weaker effect of the conditional rotations in changing the colour of the upcoming qubits.

In the end, we perform single-qubit tomography on each stroke qubit to recover the new $\phi,\, \theta$ values, which are then decoded back into hue and luminosity and used to update the colours of the segment.

In the special case where the canvas has a uniform colour and $\gamma = 1$, the result is a smooth interpolation between the brush's colour and a mixture of the canvas and brush, reminiscent of a watercolour effect. The most compelling behaviour arises for non-uniform segment colours, where quantum effects become more pronounced and produces distinctly non-classical colour blending.

\begin{figure}
\centering
\includegraphics[width=0.48\columnwidth]{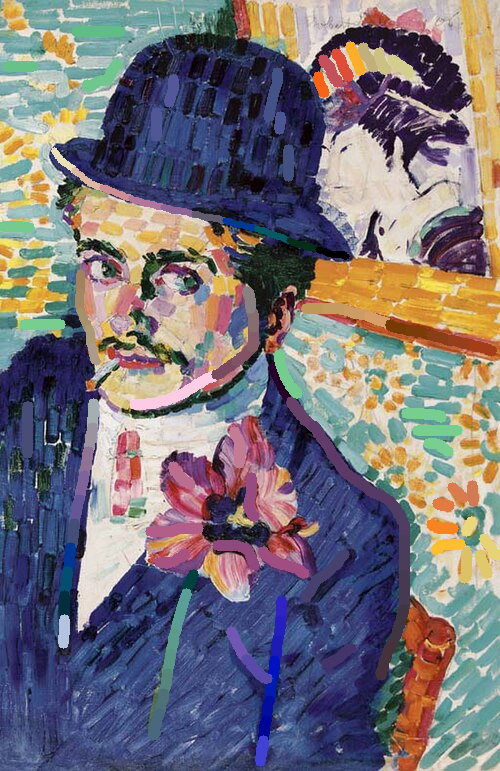}
    \includegraphics[width=0.48\columnwidth]{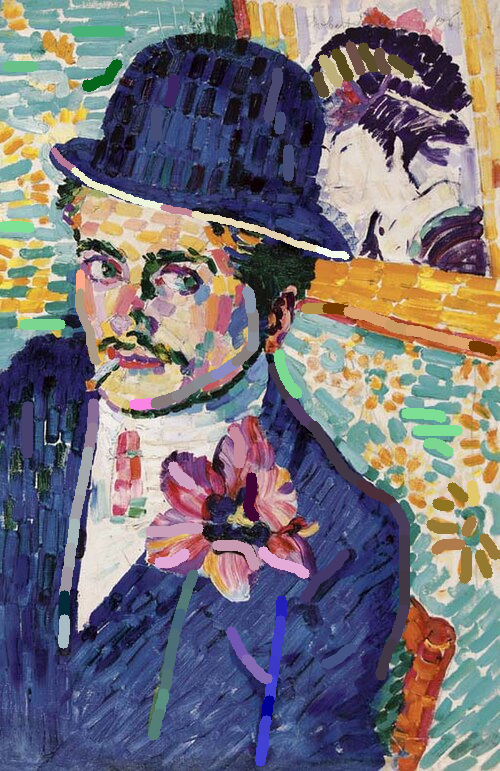}
    \caption{\textbf{Application of the Heisenbrush} on the painting \textit{L'homme à la tulipe} by Robert Delaunay, 1906~\cite{delaunay1906}. The effect is applied to recreate certain details of the composition. Left: Simulation on a noiseless backend. Right: Execution on IQM's \texttt{Sirius} device. } 
    \label{fig:fauve}
\end{figure}

\begin{figure*}[t]
\centering
\begin{tikzpicture}
\begin{yquant}
qubit {} q0;
qubit {} q1;
qubit {} q2;

box {$R_{xx}(-\Delta t)$} (q0, q1);
box {$R_{yy}(-\Delta t)$} (q0, q1);
box {$R_{zz}(-\Delta t)$} (q0, q1);

box {$R_{xx}(-\Delta t)$} (q1, q2);
box {$R_{yy}(-\Delta t)$} (q1, q2);
box {$R_{zz}(-\Delta t)$} (q1, q2);

box {$R_{xx}(-\Delta t)$} (q0, q2);
box {$R_{yy}(-\Delta t)$} (q0, q2);
box {$R_{zz}(-\Delta t)$} (q0, q2);

box {$R_{z}(\Delta t)$} q0;
box {$R_{z}(\Delta t)$} q1;
box {$R_{z}(\Delta t)$} q2;

box {$R_{x}(\Delta t)$} q0;
box {$R_{x}(\Delta t)$} q1;
box {$R_{x}(\Delta t)$} q2;

\end{yquant}
\end{tikzpicture}
\caption{\textbf{Quantum circuit of the Heisenberg model with periodic boundary conditions.} Example of a single Trotter step with $N=3$ qubits. All the qubits are measured at the end.}
\label{fig:heisenberg_circ}
\end{figure*}
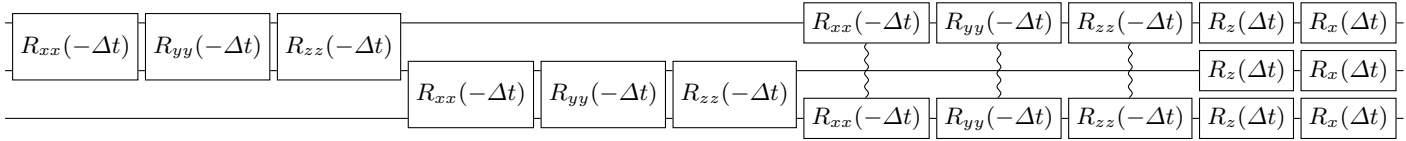

\subsection{Heisenbrush}\label{sec:heisenbrush}
\subsubsection{Motivation}
The idea behind the Heisenbrush is to generate visual patterns based on the simulated temporal evolution of quantum observables. Rather than relying on classical randomness or pre-programmed stroke behaviour, Heisenbrush draws its expressive power from the behaviour of quantum systems evolving under well-defined quantum time evolutions.
Heisenbrush allows artists to both translate the evolution of quantum mechanics into visual expression and interact with the invisible mathematical structure of quantum systems.

\subsubsection{Outcome}

Here we present an example of the application of the Heisenbrush. Fig.~\ref{fig:fauve} depicts how this brush integrates harmoniously with the vivid, expressive palette of the \textit{fauvist} painting \textit{L'homme à la tulipe} by Robert Delaunay. To create the final outcome, we used two different types of the Heisenbrush: \textit{discrete} and \textit{continuous}. In the discrete brush each stroke corresponds to a step in the time evolution, while in the continuous version we have a single stroke that is automatically split into the total number of time steps\footnote{Due to limited hardware constraints, we limit the number of steps for both brushes to a total of 10.}. 
Like an acrylic paint, Heisenbrush overwrites the canvas with its colours, making it perfect to start a new piece from a blank canvas or adding details to an existing artwork.

When applied to Delaunay's bold use of colours and abstraction, the quantum-generated strokes do not simply overlay the image but rather establish a dynamic dialogue with the original artwork. This interaction highlights how quantum computational processes can enhance and reinterpret visual art, creating a new aesthetic based on physical principles such as entanglement, superposition and non-classical evolution.

\subsubsection{Implementation}

The brush behaviour is governed by the magnetization of a one-dimensional spin-$\frac{1}{2}$ Heisenberg model with a magnetic field and periodic boundary conditions. The Hamiltonian, which describes the energy of the system, for a chain with $N$ spins reads

\begin{align}\label{hamilt_heisenberg}
    \hat{H} = \frac{1}{2}\sum_{n=1}^{N}\big( -\hat{X}_n \hat{X}_{n+1} - \hat{Y}_n \hat{Y}_{n+1} - \hat{Z}_n \hat{Z}_{n+1}  
+  \hat{X}_n + \hat{Z}_n \big),
\end{align}
where we chose the arbitrary value of $\pm 1/2$ for the exchange coupling strength and for the transverse and longitudinal fields.
$\hat{X}_n$, $\hat{Y}_n$, and $\hat{Z}_n$ denote the Pauli-$X$, Pauli-$Y$, and Pauli-$Z$ operators acting on the $n$-the qubit:

\begin{align}
\hat{X} = 
\begin{pmatrix}
0 & 1 \\
1 & 0
\end{pmatrix}, \ \ \ 
\hat{Y}= 
\begin{pmatrix}
0 & -i \\
i & 0
\end{pmatrix}, \ \ \ 
\hat{Z} = 
\begin{pmatrix}
1 & 0 \\
0 & -1
\end{pmatrix}.
\end{align}
The number of qubits grows linearly with the radius of the brush and is capped at 10 qubits.  

Immediate implementation of an arbitrary Hamiltonian is never possible on quantum computers. Instead, a standard technique to simulate the time evolution is via first-order Trotter-Suzuki decomposition~\cite{suzuki1976generalized,trotter1959product}:
\begin{align}
e^{-i\hat{H}t} \approx \left(\prod_{n} e^{-i\hat{H}_n \Delta t} \right)^k,
\end{align}
where $\hat{H} = \sum_n \hat{H}_n$ is decomposed into local terms acting on neighboring qubits, and $k$ is the number of Trotter steps. In this work we fix the time step at $\Delta t=0.1$.

The quantum circuit for a single Trotter step, see Fig.~\ref{fig:heisenberg_circ}, consists of the following operations:

\begin{itemize}
    \item \textbf{Two-qubit interaction terms}:
    For each neighboring pair $(n, n+1)$, we apply three 
    2-qubits gates that correspond to the exponentials of the Pauli interaction terms
        \begin{align}
       & R_{xx}(-\Delta t)=e^{i \Delta t \hat{X}_n \hat{X}_{n+1} /2},  \label{eq:rxx}\\
        &R_{yy}(-\Delta t)=e^{i\Delta t \hat{Y}_n \hat{Y}_{n+1} /2}, \\
       & R_{zz}(-\Delta t)=e^{i\Delta t \hat{Z}_n \hat{Z}_{n+1} /2}.
 \end{align}

    \item \textbf{Single-qubit field terms}:
For each qubit $n$, the local transverse and longitudinal fields are implemented as:
        \begin{align}
        &R_{x}( \Delta t)=e^{-i  \Delta t \hat{X}_n /2}, \\
        &R_{z}(\Delta t)= e^{-i\Delta t \hat{Z}_n /2} . \label{eq:rz}
        \end{align}
\end{itemize}

The trotterized Hamiltonian is then used to simulate the step-by-step time evolution of a uniform initial quantum state, 
\begin{equation}
\ket{\psi(t=0)} = \prod_{i=1}^{N} \ket{(\phi,\theta)},
\end{equation}
where the angles $(\phi,\theta)$ are obtained from a user-selected colour as described in App.~\ref{app:hsl}.

At every step in the evolution, we measure the mean expectation value of the magnetization $M_z$

\begin{equation}
    \langle M_z \rangle(t) = \frac{1}{N} \sum_{i=1}^{N} \bra{\psi(t)} Z_i  \ket{\psi(t)},
\end{equation}
and use it to compute the HSL values of the $n^{th}$ segment according to
\begin{equation}
    V_n = \gamma [V_{u} + \langle M_z \rangle(n \Delta t) \mod 1] + (1-\gamma) V_{u},
\end{equation}
where $V$ corresponds to H, S and L values and the subscript $u$ to the user-selected colour. The strength $\gamma$ controls how much the quantum-evolved colours influence the selected colour. At low strength, the original colour dominates; at high strength, the quantum behaviour takes over, leading to colour transformations.

We provide two versions of the Heisenbrush. The \textit{continuous} brush segments a single stroke by the number of time steps. For the \textit{discrete} brush, each stroke of the user corresponds to a single time step and thus a single colour, related to the expectation value of the magnetization at that time. In both cases, we limit the number of time steps to 10 to avoid long execution times.

\subsection{Smudge}\label{sec:qdamp}

\subsubsection{Motivation}

The Smudge brush explores the mechanisms of information erasure in quantum systems. As the brush glides across the canvas, it first attempts to erase the information about the colour, but in doing so, it creates a source of new entangled colours that are transferred to the remainder of the brush stroke. 

\subsubsection{Outcome}

\begin{figure}[]
    \centering
    \includegraphics[width=0.48\columnwidth]{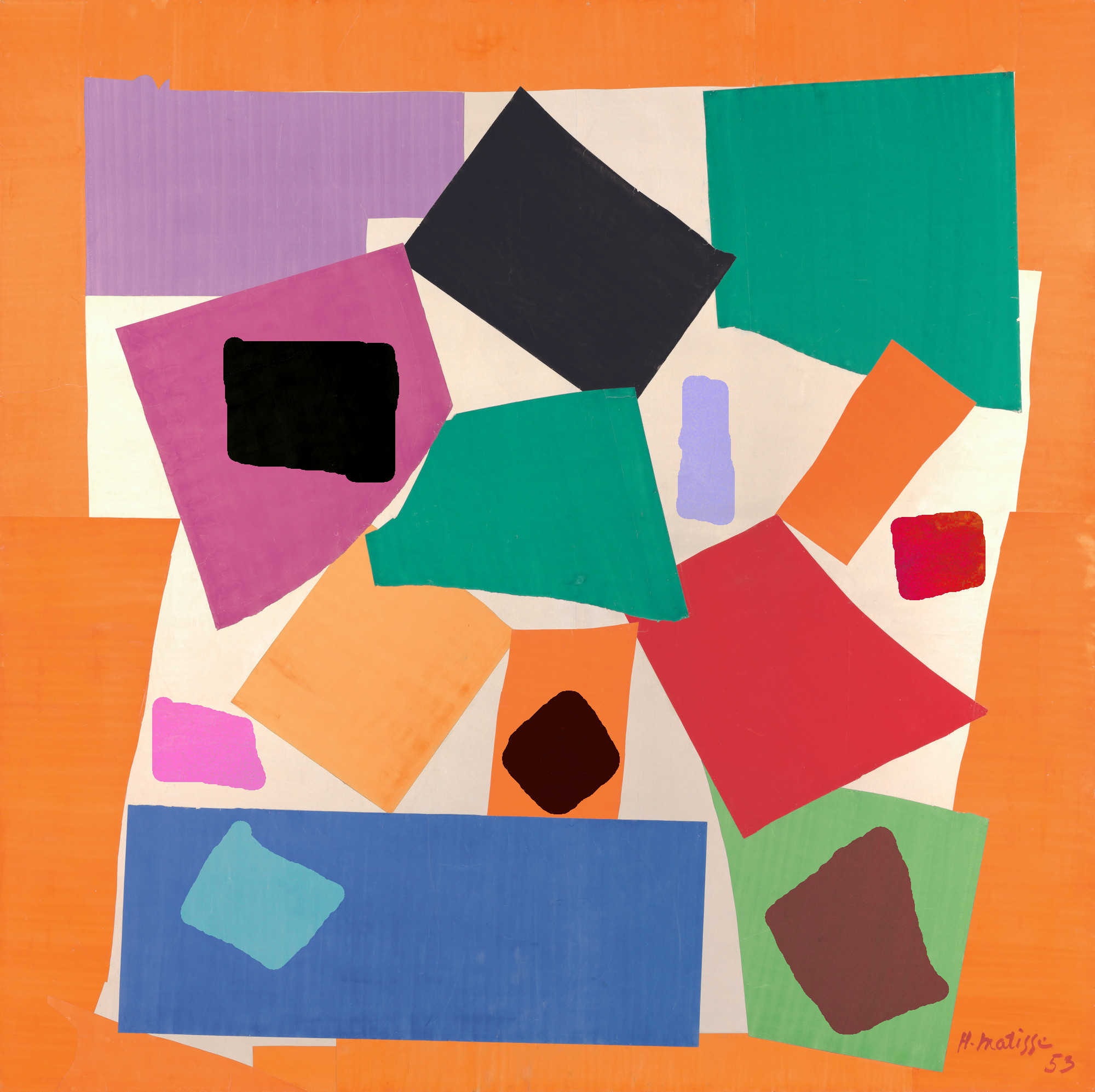}
    \includegraphics[width=0.48\columnwidth]{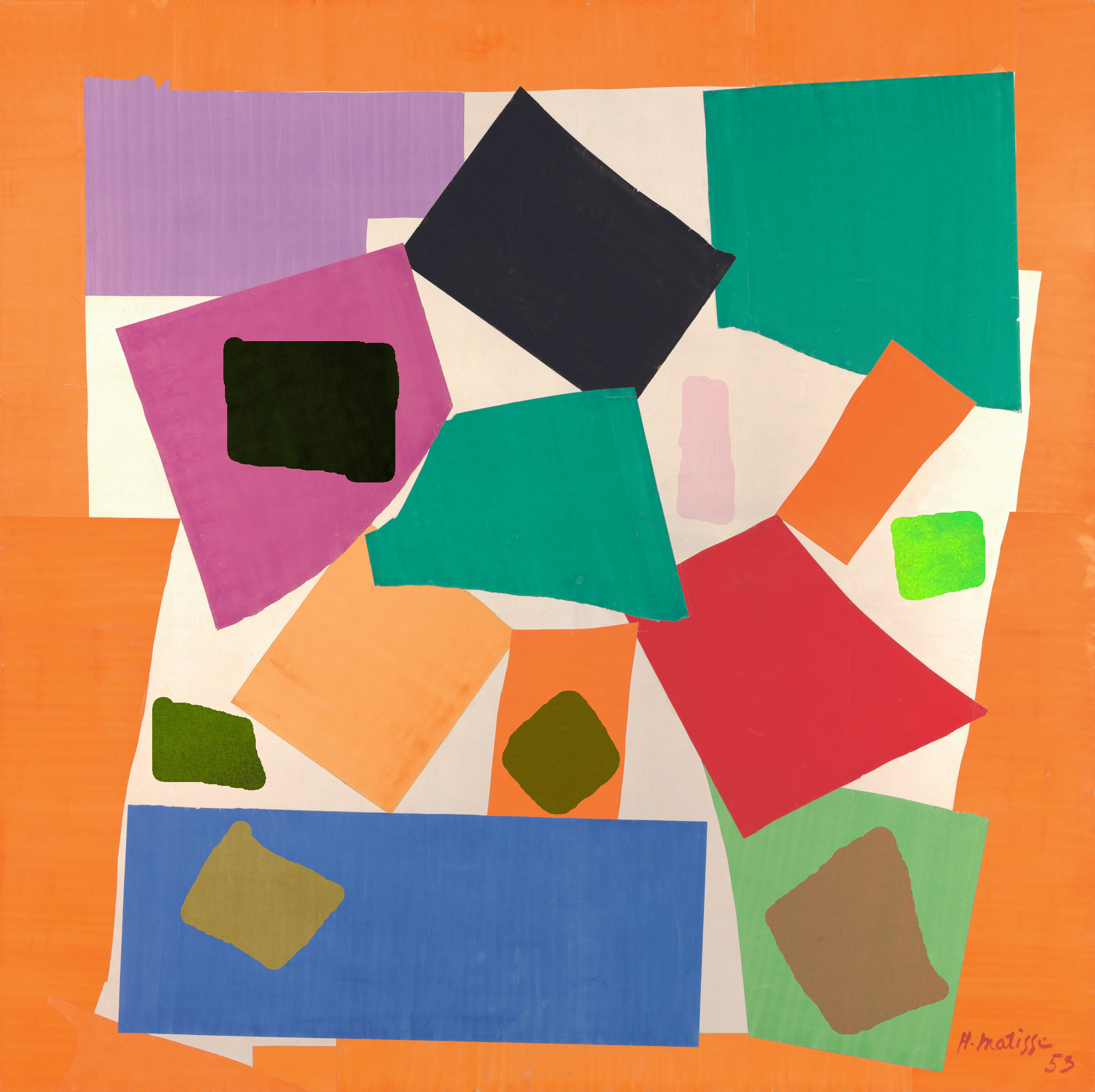}
    \caption{\textbf{Application of the Smudge} on the artwork \emph{L'escargot} by Henri Matisse, 1953~\cite{mattisse1953}. Left: Simulation on a noiseless backend. Right: Execution on IQM's \texttt{Sirius} device.}
    \label{fig:snail}
\end{figure}

In our interpretation of Matisse's \emph{L'escargot}, Fig.~\ref{fig:snail}, we use a single run of Smudge (composed of 7 strokes) to create shapes with new colours.
Each stroke was applied on a different canvas colour, leading to a highly entangled state that maximizes the colour expressivity and variability. 
We applied a maximum strength and did not invert luminosity, identifiable by the black shape on the top left part instead of a white one.  

\begin{figure}
\centering
\begin{tikzpicture}
  \begin{yquant}
    cbit {Control} c;
    qubit {$\ket{(\phi_{1},\theta_{1})}$} q;
    qubit {\ldots} n;
    qubit {$\ket{(\phi_{N},\theta_{N})}$} q[+1];
    qubit {$\ket{0}$} a;

    x q[0] | c;
    box {R$_y(\gamma)$} a | q[0];
    cnot q[0] | a;
    x q[0] | c;
    x q[1] | c;
    box {R$_y(\gamma)$} a | q[1];
    cnot q[1] | a;
    x q[1] | c;
    
\end{yquant}
\end{tikzpicture}
\caption{\textbf{Implementation of the Smudge brush.} A shared ancilla qubit applies the amplitude damping channel to every qubit on the canvas, controlled by the user input $\gamma$. The user can also decide if the effect will darken or brighten the first colour through a controlled $X$ gate. All qubits are measured at the end.}
    \label{fig:qdamp_circ}
\end{figure}
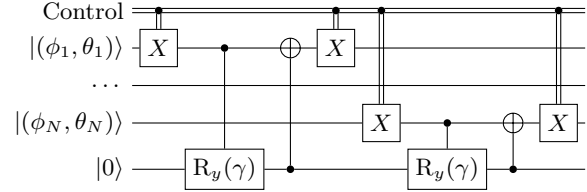

Again, we observe clear colour differences between the simulated and real runs for almost all new shapes. Possibly due to the presence of dephasing noise, the real run tends towards darker colours (dark browns and dark greens) but still provides us with a revelation of bright green.

\subsubsection{Implementation}

The user starts by creating $N$ separate strokes on the canvas. For each stroke, we compute the average hue and luminosity and map it to the single-qubit state $\ket{(\phi,\theta)}$, see App.~\ref{app:hsl}. We also initialize an ancilla qubit in the $\ket{0}$ state, for a total of $N+1$ qubits.
The ancilla qubit then interacts sequentially with all the canvas's qubits as shown in Fig.~\ref{fig:qdamp_circ}.

The first step of the circuit implements an amplitude damping or pumping channel~\cite{chen_2021}, respectively for $\text{Control}=0,1$. This quantum channel captures the non-unitary, non-Hermitian effect of dephasing: it collapses the first qubit state toward the states $\ket{0}$ or $\ket{1}$, independently of its initial state. The strength of this damping/pumping is controlled by a parameter $\gamma$, where $\gamma = \pi$ corresponds to complete collapse and $\gamma = 0$ to no change.
In this process, information carried in the first qubit is transferred to the ancilla qubit which should be reset at the end of a simulation of damping or pumping, thus erasing the information. 

However, in our implementation, the ancilla qubit is never re-initialized and the damping/pumping channel is applied to the remaining canvas qubits in a sequence.
As a result, only the first qubit undergoes true amplitude damping/pumping. The remaining qubits become entangled through the shared ancilla, forming a complex many-body state. The overall effect is not uniform decoherence, but rather an uneven quantum cascade, where only the tip of the brushstroke experiences pure dissipation.

After the circuit evolution, we perform single-qubit tomography on each stroke qubit and reconstruct the new average hue and luminosity as detailed in App.~\ref{app:hsl}. 

In the scaling limit $N\gg 1$ and in the special case where all segments have approximately the same colour, most qubits will converge to the $\ket{\phi=0,\theta=\pi/2}$ and $\ket{\phi=\pi,\theta=\pi/2}$ state, respectively for $\gamma\approx 0$ and $\gamma \approx 1$. This implies the existence of a transition $\gamma$, where we empirically observe the most colour variability between segments.

\subsection{Collage}
\label{sec:ctrlq}
\subsubsection{Motivation}

The motivation behind the Collage brush is to visualize one of the most important intrinsic limitations imposed by quantum mechanics: the no-cloning theorem. It states that it is impossible to make an arbitrarily faithful copy of an unknown quantum state without completely collapsing the original. When attempting to copy part of an image, a balance must be struck: increasing the fidelity of the new copy necessarily sacrifices information in the original, and vice versa.

\subsubsection{Outcome}

\begin{figure}
    \centering
    \includegraphics[width=0.47\columnwidth]{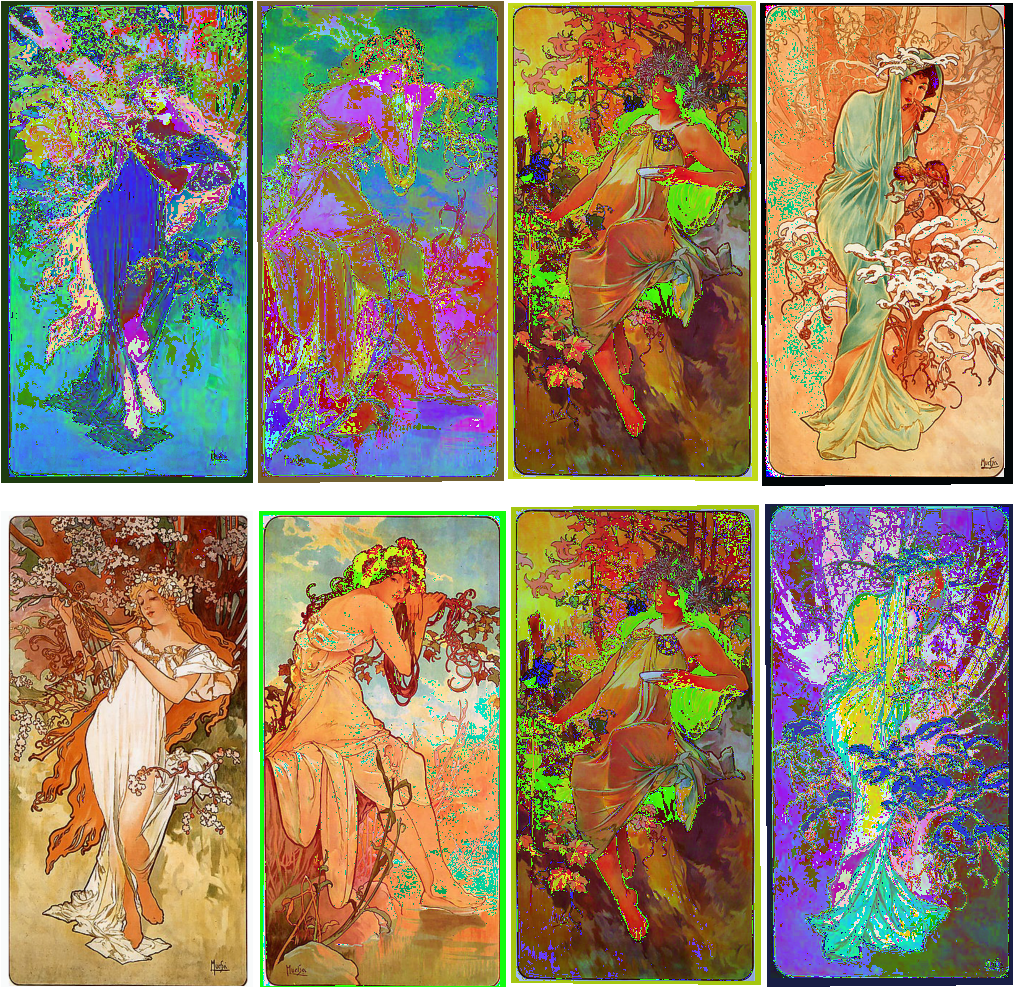}
    \hspace{0.03\columnwidth}
     \includegraphics[width=0.47\columnwidth]{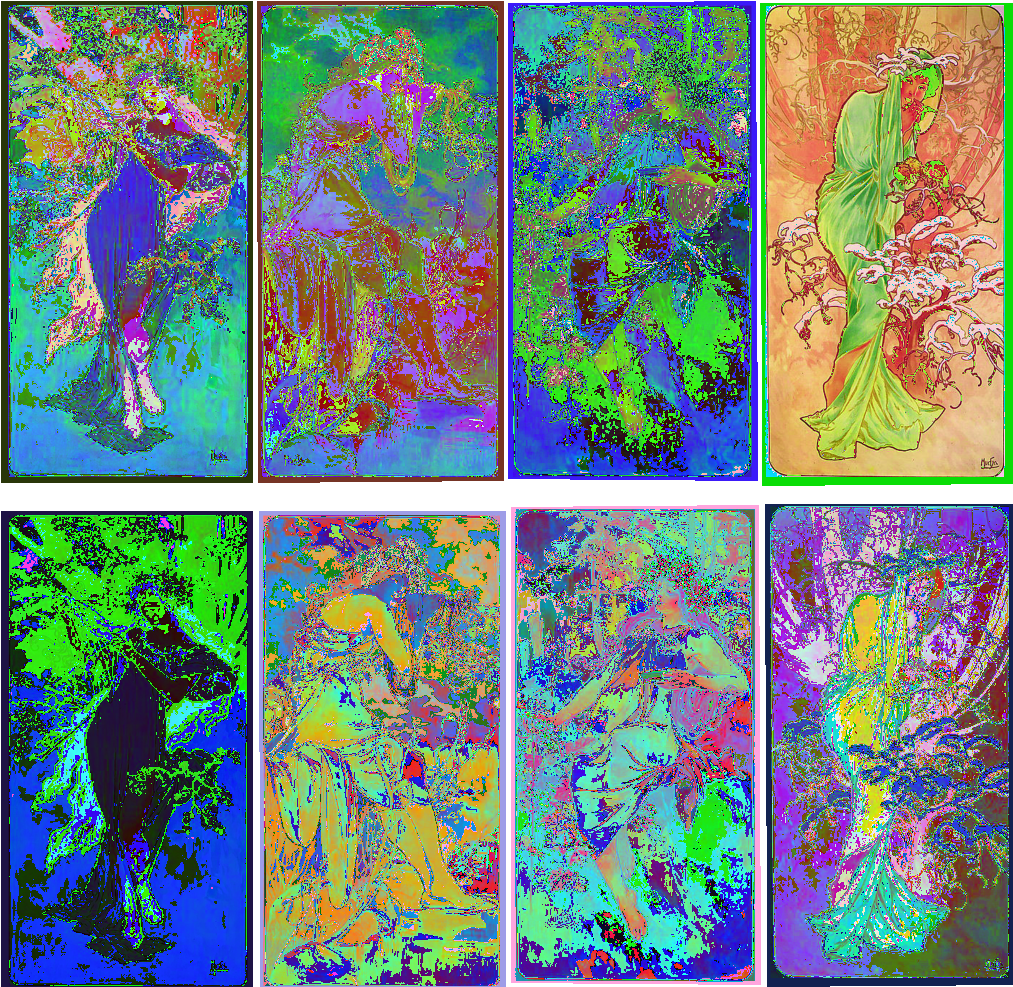}
    \caption{\textbf{Application of the Collage brush.} Inspired by serial art, we create four copies of \emph{Les Saisons} by Alfons Maria Mucha~\cite{alfons}, 1896, with various degrees of fidelity. Left: Simulation on a noiseless backend. Right: Execution on IQM's \texttt{Sirius} device.}
    \label{fig:ctrlq}
\end{figure}

In Fig.~\ref{fig:ctrlq}, we demonstrate how the Collage brush can be used to create a serial art motif from a sequence of artworks, in particular, \emph{Les Saisons} by Alfons Maria Mucha~\cite{alfons}, 1896. From top left to top right, we apply the Collage effect to copy one season at a time to the bottom with increasing degrees of strength (interpolating from high to low fidelity of the copied image). Notice how the defining lines of artwork remain largely visible across all outputs as the Collage brush primarily alters internal colour correlations rather than structural features. The advantage of this brush is the capacity to modify large regions of the canvas using a single shallow circuit.

Due to an exponentiation step discussed below, the Collage brush is very sensitive to noise. This is clear from the outcome of the \texttt{Sirius} device, see Fig.~\ref{fig:ctrlq}-right, where only the colours of the last muse resemble the simulated case.

\subsubsection{Implementation}

Unlike other brushes, the Collage brush does not operate through a click-and-drag movement. Instead, the user selects a lasso-defined \emph{copy} region and a target \emph{paste} location on the canvas.

The brush encodes the RGB content of the copy region into a \emph{single} qubit, denoted $C$, in the following way. First, we extract the RGB values of all pixels in the selected region, forming a rectangular matrix of width 3. We then perform a singular value decomposition (SVD) to obtain three singular values $\vec{S} = [S_0, S_1, S_2]$ and their corresponding singular vectors.

The copy qubit is initialized in the state
\begin{equation}
\ket{\psi}_C = R_z(\phi) R_y(\theta) \ket{0},
\end{equation}
where the spherical angles are determined by:
\begin{equation}
\label{eq:bloch_angles}
\tan \phi = \frac{\log S_1}{\log S_0}, 
\qquad
\tan \theta = \frac{\sqrt{(\log S_0)^2 + (\log S_1)^2}}{\log S_2}.
\end{equation}
This effectively orients the qubit on the Bloch sphere along the direction of $\log \vec{S}$. We use logarithms to mitigate the exponential decay in the singular spectrum, making sure that every image is uniquely encoded. 

\begin{figure}[]
\centering
\begin{tikzpicture}
  \begin{yquant}
    qubit {$\ket{\psi}_C$} q;
    qubit {$\ket{0}_A$} q[+1];
    qubit {$\ket{0}_P$} q[+1];

    box {
    U$_P$} (q[1, 2]) ;
    cnot q[2] | q[0];
    cnot q[1] | q[0];
    cnot q[0] | q[2];
    cnot q[0] | q[1];

  \end{yquant}
\end{tikzpicture}
\caption{\textbf{Universal asymmetric quantum cloning circuit implementation.} The U$_P$ transformation prepares the cloned qubit (P) and a necessary ancilla (A); the four CNOT gates perform the cloning step properly speaking. The ancilla may then be discarded. }
    \label{fig:UAQC}
\end{figure}
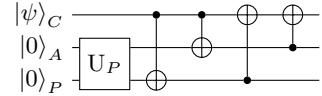

Next, we apply a universal asymmetric quantum cloning (UAQC) protocol \cite{cerf,Iulia_2003} to attempt an approximate $1 \rightarrow 2$ cloning of the state $\ket{\psi}_C$ onto a second qubit, $P$. The ideal cloning transformation,
\begin{equation}
U\,\ket{\psi}_C \ket{0}_P = \ket{\psi}_C \ket{\psi}_P,
\end{equation}
is forbidden by the no-cloning theorem, but UAQC allows imperfect cloning. Its circuit implementation is shown in Fig.~\ref{fig:UAQC}.
It uses a single ancilla qubit (A), prepared in a shared state with $P$ and involves two steps. The first preparation step entangles A and P using the transformation $U_P$ such that:
\begin{equation}
    U_P\lvert00\rangle=\sqrt{\frac{s_0 + s_1}{2}}\lvert 00 \rangle + \sqrt{\frac{1 - s_0}{2}}\lvert 01\rangle + \sqrt{\frac{1 - s_1}{2}}\lvert 11 \rangle,
\end{equation}
where $s_{0,1}$ are two independent parameters that control the fidelity of the clones. The second cloning step is made of four CNOT gates and involves all three qubits. The resulting density matrices for the copy and cloned qubit are
\begin{equation}
\label{eq:uaqc_densities}
\begin{aligned}
\rho_C &= s_0\,\ket{\psi}\bra{\psi}_C + \frac{1 - s_1}{2} I, \\[6pt]
\rho_P &= s_1\,\ket{\psi}\bra{\psi}_P + \frac{1 - s_0}{2} I,
\end{aligned}
\end{equation}
where $I$ is the identity matrix.
Note that $s_{0,1}$ cannot be both unity simultaneously and must satisfy
\begin{equation}
s_0^2 + s_1^2 + s_0 s_1 - s_0 - s_1 \le 0.
\end{equation}
In our implementation, $s_0$ is a user-defined control parameter; $s_1$ is then chosen to maximize the fidelity of the paste output $\rho_P$. The optimal symmetric setting is obtained for $s_0 = s_1 = 2/3$.

After cloning the qubit, we perform single-qubit state tomography on both $C$ and $P$ to extract the vector of Pauli expectation values
\begin{equation}
\vec{E}_q = 
\left[
\langle     \hat{X}_q \rangle,
\langle \hat{Y}_q \rangle,
\langle \hat{Z}_q \rangle
\right], \quad q \in \{C, P\}.
\end{equation}

We map each $\vec{E}_q$ back to a new set of singular values $\vec{S}_q = [S_{0,q}, S_{1,q}, S_{2,q}]$ using
\begin{equation}
\label{eq:singular_mapping}
\vec{S}_q = \lvert \vec{E}_q \rvert \cdot \exp\!\left( \frac{\lvert \log \vec{S} \rvert}{\lvert \vec{E}_q \rvert} \cdot \vec{E}_q \right)
+ \left(1 - \lvert \vec{E}_q \rvert\right) \cdot \overline{S},
\end{equation}
where $\overline{S} = \frac{1}{3}(S_0 + S_1 + S_2)$ is the average of the original singular values. This mapping satisfies the following properties:
\begin{itemize}
    \item If $\vec{E}_q = \log \vec{S} / \lvert \log \vec{S} \rvert$, then $\vec{S}_q = \vec{S}$; recovering the original values in the case that the cloning is perfect.
    \item If $\vec{E}_q = [0,0,0]$ (a maximally mixed state), then $\vec{S}_q = [\overline{S}, \overline{S}, \overline{S}]$ and all the information about the colours is lost.
\end{itemize}

Finally, we reconstruct image patches using the new singular values $S_C$ and $S_P$ along with the original singular vectors.

\section{User Application}
\label{sec:usage}

To incentivize the exploration of the brushes by artists, we introduce the associated application \textbf{Quantum Brush} \cite{quantumbrush_repo} as an open-source micro-scale software. This multi-platform application is built with Java, Processing \cite{processing_org} external library and Python for the quantum algorithm backend.

The interaction between Quantum Brush and users is similar to a typical image modification software, see Fig.~\ref{fig:app}. Users can import the image on the \textit{canvas} and from the control panel window select the chosen brush and parameters. They then draw strokes on the specific region they want to apply quantum effects and submit them to a special \emph{Stroke Manager} window. 

The Stroke Manager supports running each brush concurrently. That is, the user workflow will not be affected by execution on quantum devices, which can consume considerable time between submission and execution. To do so, when a stroke is added to the Stroke Manager, a snapshot of the canvas is saved and used as the reference image for the quantum effect. After executing the stroke, the updated image is compared to the snapshot and only the modified pixels will be pasted onto the current canvas. 
This snapshot approach allows the user to re-execute the same strokes as much as they want until they paste the best outcome, thus providing creative exploration of the stochasticity of quantum computing. 

Furthermore, the merit of Quantum Brush is that any creative technologist could develop novel quantum visual effects. All new effects can be managed locally and automatically added to the app without separate import statements. This shortens the user workflow and assists users in collaborating with the program on a deeper level, inviting further contribution to open-source quantum software projects.

\begin{figure*}[ht]
    \centering
    \includegraphics[width=0.55\linewidth]{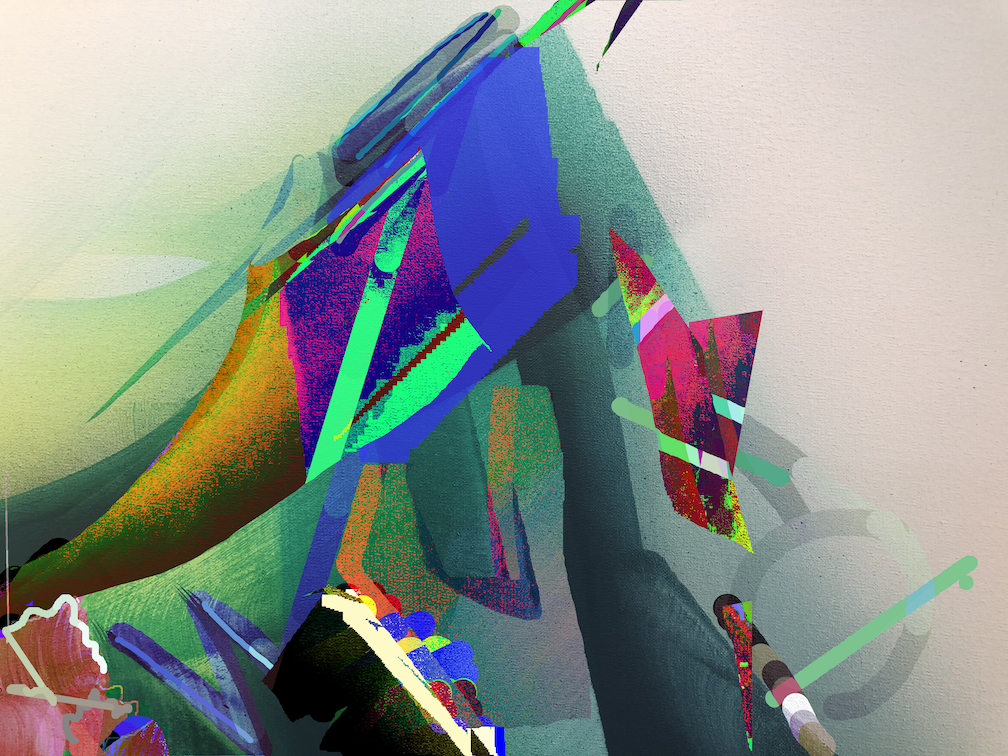}
    \caption{\textbf{Composition No.~1} by Roman Lipski. All the available brushes were used in this composition as well as a simple classical fill-in brush. The starting point was the photography of a painted canvas.}
    \label{fig:lipski}
\end{figure*}

\section{Artist's Experience}

\label{sec:artwork}

The integration of quantum processes into creative practice offers novel forms of interaction with colour, structure, and composition. To investigate these possibilities, one of the authors (R.L.) employed the Quantum Brush in the creation of novel digital artworks. Fig.~\ref{fig:lipski} illustrates an example produced by combination of several brush effects.

A central feature of this exploration was the unfamiliarity of the medium. Whereas classical painting techniques rely on well-established rules, such as mixing pigments on a palette, the Quantum Brush introduced behaviours and transformations that did not follow these conventions. These novel mechanisms produced unexpected results, challenging established habits and prompting new creative directions. The experience was described as a `dialogue' with the tool, in which its responses continually suggested variations and approaches that would not have been anticipated using conventional techniques.

Distinct roles were observed for individual brushes. The Collage effect enabled large-scale transformations while preserving global structure, accelerating the early stages of composition. The other brushes were applied more selectively to enrich fine details and textures, with transparency settings allowing for subtle layering. Iterating between coarse global transformations and fine refinements produced compositions of increasing complexity and balance, supporting a workflow that oscillated between experimentation and control. To support this workflow, a pressure-sensitive tablet was necessary, as mouse input did not provide the required level of fine detail.

Beyond technical use, the Quantum Brush also influenced considerations of colour and composition. Traditionally, the symbolic significance of colour imposed constraints that could restrict creative decisions. The novel transformations produced by the Quantum Brush allowed a freer exploration of compositional possibilities. Each digital image became an experiment in its own right, which could remain abstract or later be adapted onto canvas. The tool therefore functioned both as a support for existing ideas and as a generator of new creative directions, with potential applications extending into immersive media such as virtual reality.

\section{Conclusions}
\label{sec:discussion}
In this work, we introduced an open-source tool that invites artists to explore the quantum world through a familiar and intuitive interface: the brush. By merging the abstract mathematics of quantum mechanics with the tangible gesture of a stroke on a canvas, we aim to make the invisible mechanisms and beauty of quantum processes visible and provide a personal touch.

We presented four distinct quantum brushes, Aquarela (Sec.~\ref{sec:Aquarela}), Heisenbrush
(Sec.~\ref{sec:heisenbrush}), Smudge (Sec.~\ref{sec:qdamp}) and Collage (Sec.~\ref{sec:ctrlq}). Each of these brushes is rooted in real physical principles and can be implemented in today's quantum hardware. Every brush carries its individual aesthetic signature, leading to artworks that capture the underlying quantum effect in visual form. The experiments we carried out demonstrated the rich potential of the quantum brush at the intersection of quantum theory and artistic intuition. In particular, the unique dynamics of the quantum processes led to unexpected results that can create new artistic inspiration and expression. We believe that such quantum dynamics provide unique perspectives and therefore an artistic advantage over machine learning approaches, whose overdependence on vast training datasets leads to flattening of artistic expression~\cite{chayka2024filterworld}.   

Though we focused on examples where the quantum brushes serve to modify existing paintings, they can also be used on a blank canvas to create completely new artwork. The different brushes, by themselves or in combination, offer a new, so far unexplored way to create visual art. Although the digital images are already very appealing, an exciting prospect is to bring these images to real canvases, which would give back the human touch, as also explored in \cite{crippa2025quantumcomputinginspiredpaintings}.

We also tested all brushes on IQM’s \texttt{Sirius} device and deliberately chose to omit error mitigation, enabling thus hardware noise in combination with quantum noise as part of the underlying quantum aesthetics. This approach highlights the interplay between computational and physical limitations and artistic expression, demonstrating what is both computationally and artistically possible with quantum devices even now.

As part of future work, we aim to develop new quantum effects and evaluate them on different QPUs. We are also interested in providing interactive colour pallettes for each brush, as well as differently formed brushes allowing artists to anticipate the novel ways of interacting with our quantum-native colour space. 
In this study, we focused on modifying colour spaces, but similar algorithms or physical models could also extend painting into areas like structure and composition, where virtual reality could play an important role. Thus we consider the present work as an initial step only towards a powerful tool which employs quantum-mechanical principles to create novel, unseen visual art.

Quantum Brush is freely available as an open-source project~\cite{quantumbrush_repo}. We invite artists, scientists, and dreamers alike to contribute their own brushes, algorithms, and ideas. Together, we can continue to shape the emerging movement of quantum aesthetics, where art and quantum physics coalesce as collaborators.

\begin{acknowledgments}
The authors would like to thank Declan Colquitt, Harry Kumar, Marcel Pfaffhauser, Birgit Ostermeier and Walter Riess for discussions.

This work is supported with funds from the Ministry of Science, Research and Culture of the State of Brandenburg within the Centre for Quantum Technologies and Applications (CQTA). 
\begin{center}
    \includegraphics[width = 0.1\textwidth]{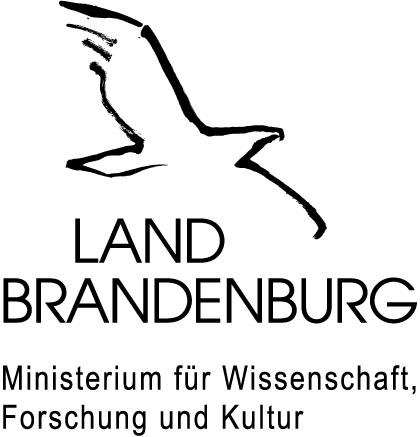}
\end{center}
This work is funded by the European Union’s Horizon Europe Framework Programme (HORIZON) under the ERA Chair scheme with grant agreement no. 101087126.
We acknowledge the use of IQM quantum services for this work.
\end{acknowledgments}

\appendix
\section{HSL colours and qubits\label{app:hsl}}

\begin{figure}[h]
    \centering
    \includegraphics[width=0.44\columnwidth]{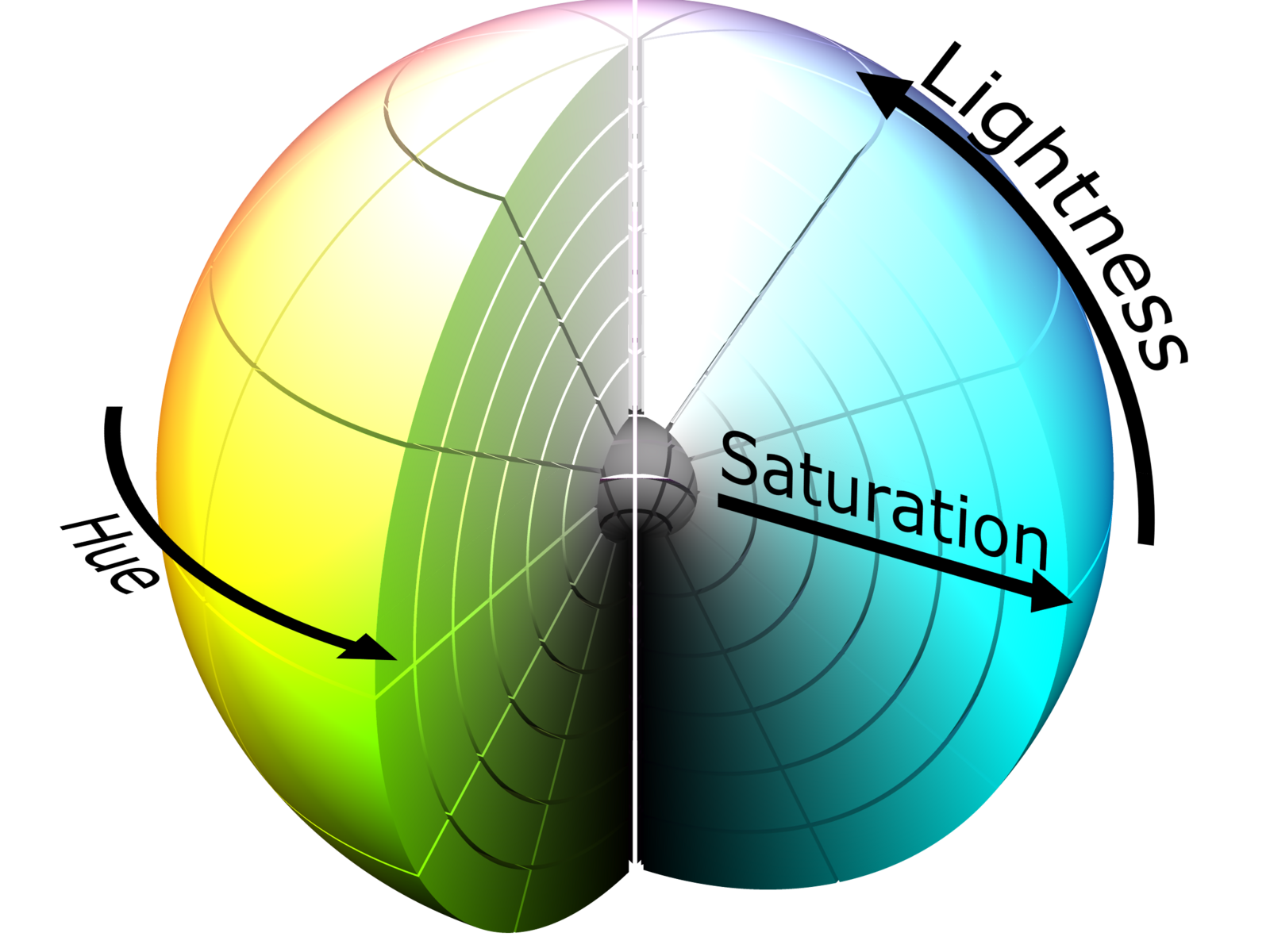}
    \includegraphics[width=0.48\columnwidth]{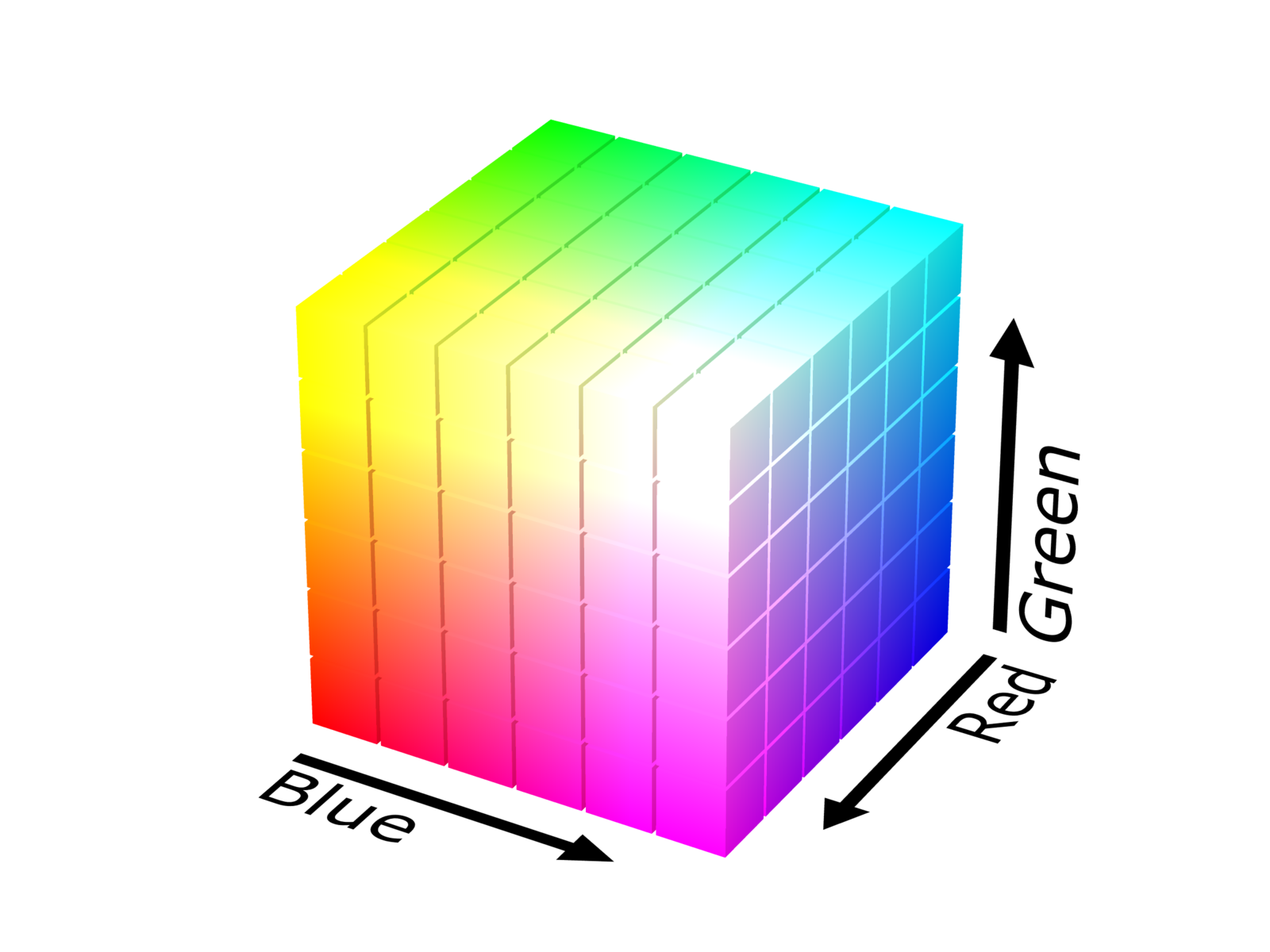}
    \caption{\textbf{Comparison between HSL and RGB colour mapping schemes.} Left: HSL-based colour mapping in spherical coordinates. Right: RGB-based colour mapping in cartesian coordinates.}
    \label{fig:hsl_vs_rgb}
\end{figure}

In this work, we frequently encode colour information directly into single-qubit states, rather than representing it via amplitudes in Fock space. This design choice aligns well with the constraints of NISQ devices, as colour transformations can then be executed using only single- and two-qubit gates.
However, a direct mapping from RGB (Red, Green, Blue) to qubit states is non-trivial. A core challenge lies in the fact that summing two RGB colours can produce a third colour outside the representable RGB gamut, leading to unphysical results or saturation effects when quantum operations are applied. To address this, we adopted the HSL (hue, saturation, luminosity) colour model, illustrated in Fig.~\ref{fig:hsl_vs_rgb}-left, instead of RGB mapping, Fig.~\ref{fig:hsl_vs_rgb}-right.

The HSL colour space maps naturally onto spherical coordinates. Hue (H) is an angular quantity on the colour wheel and aligns well with the azimuthal angle $\phi$ on the Bloch sphere. Luminosity (L) corresponds to the polar angle $\theta$, while saturation (S) can be viewed as a radial component. This geometric correspondence makes it natural to map HSL colours onto the Bloch sphere, with pure colours lying on the surface and desaturated colours moving inward.

Nonetheless, this mapping is not without limitations. To represent a fully general HSL colour with all three components, one would need to prepare a mixed quantum state—particularly when dealing with desaturated (low-S) colours. This introduces additional quantum operations and noise, which is undesirable for NISQ devices. Importantly, one cannot simply `stretch' the Bloch vector to match full saturation, as a noisy quantum dynamics most often shrinks the Bloch vector. This would result in unintended graying or colour loss. To avoid this, we chose to encode only the hue and luminosity components into the qubit state, leaving the saturation unchanged.

The mapping from HL to a pure qubit state is given by the following operation:
\begin{equation}
    \ket{(\phi,\theta)} = R_z(\phi)R_y(\theta)\ket{0},
\end{equation}
where
\begin{equation}
    \phi = 2\pi \, \text{H}, \qquad \theta = \pi \, \text{L},
\end{equation}
and (H,L) are between 0 and 1.
This prepares a pure state whose Bloch vector direction corresponds to the desired hue and brightness.

In the case of the Smudge and Aquarela brushes, where we need to compute the mean hue and luminosity over a region. While L is a linear quantity and can be averaged arithmetically, H is angular and must be averaged using the circular mean:
\begin{equation}
\bar{\text{H}} = \arg\left( \sum_{k} e^{i 2\pi \text{H}_k} \right),
\end{equation}
ensuring continuity across the hue boundary at 0/1. This avoids artifacts where hues wrap around the colour circle.

To reconstruct the colour of a qubit after applying some quantum process, we performed single-qubit tomography to extract the expectation values of the Pauli operators $ \langle X \rangle, \langle Y \rangle, \langle Z \rangle $.

From these, the colour angles can be retrieved using:
\begin{equation}
    \label{eq:bloch_reconstruction}
    \tan \phi \;=\; \frac{\langle Y \rangle}{\langle X \rangle}, 
    \qquad
    \tan \theta \;=\; \frac{\sqrt{\langle X \rangle^2 + \langle Y \rangle^2}}{\langle Z \rangle}.
\end{equation}
These angles can then be mapped back to hue and luminosity. The saturation value, which remained unchanged during the encoding, is simply reused to reconstruct the full HSL color. 
In the case of remapping the colour of a region, we uniformly shift the H and L values to the new mean, preserving the angular nature of H and cropping any overflow values in the luminosity. 

This approach allows us to represent colour in a quantum-native, NISQ-compatible manner, while preserving artistic control and interpretability.

\bibliography{bibliography}

\end{document}